%% file: main.tex
\documentclass[sigconf]{acmart}

\AtBeginDocument{%
  }

\copyrightyear{2026}
\acmYear{2026}
\setcopyright{cc}
\setcctype{by}
\acmConference[KDD 2026] {Proceedings of the 32nd ACM SIGKDD Conference on Knowledge Discovery and Data Mining V.2}{August 9--13, 2026}{Jeju Island, Republic of Korea.}
\acmBooktitle{Proceedings of the 32nd ACM SIGKDD Conference on Knowledge Discovery and Data Mining V.2 (KDD 2026), August 9--13, 2026, Jeju Island, Republic of Korea}
\acmISBN{979-8-4007-2259-2/2026/08}
\acmDOI{10.1145/3770855.3818146}

\usepackage{hyperref}
\usepackage{algorithm2e}
\usepackage{multirow}
\usepackage{balance}
\usepackage{amsfonts}
\usepackage{graphicx}  

\usepackage{amssymb}

\usepackage{colortbl}
\usepackage{pifont}
\usepackage{makecell}
\usepackage{enumitem} 
\usepackage{colortbl}

\begin{document}
\input{dfn}

\title{SAGEO Arena: A Realistic Environment for Evaluating Search-Augmented Generative Engine Optimization}


\author{Sunghwan Kim}
\orcid{0009-0003-9514-812X}
\affiliation{%
  \department{Department of Artificial Intelligence}
  \institution{Yonsei University}
  \city{Seoul}
  \country{Republic of Korea}
}
\email{happysnail06@yonsei.ac.kr}

\author{Wooseok Jeong}
\orcid{0009-0007-1931-4076}
\affiliation{%
  \department{Department of Computer Science and Engineering}
  \institution{Konkuk University}
  \city{Seoul}
  \country{Republic of Korea}
}
\email{jws010825@konkuk.ac.kr}

\author{Serin Kim}
\orcid{0009-0009-2413-1121}
\affiliation{%
  \department{Department of Artificial Intelligence}
  \institution{Yonsei University}
  \city{Seoul}
  \country{Republic of Korea}
}
\email{kimserin@yonsei.ac.kr}

\author{Sangam Lee}
\orcid{0009-0000-2479-7606}
\affiliation{%
  \department{Department of Artificial Intelligence}
  \institution{Yonsei University}
  \city{Seoul}
  \country{Republic of Korea}
}
\email{salee@yonsei.ac.kr}

\author{Dongha Lee}
\orcid{0000-0003-2173-3476}
\authornote{Corresponding author}
\affiliation{%
  \department{Department of Artificial Intelligence}
  \institution{Yonsei University}
  \city{Seoul}
  \country{Republic of Korea}
}
\email{donalee@yonsei.ac.kr}

\renewcommand{\shortauthors}{Sunghwan Kim, Wooseok Jeong, Serin Kim, Sangam Lee, \& Dongha Lee}

\begin{abstract}
\input{MainText/000_abstract}

\end{abstract}

\maketitle

\newcommand\kddavailabilityurl{https://doi.org/10.5281/zenodo.20423531}
\ifdefempty{\kddavailabilityurl}{}{
\begingroup\small\noindent\raggedright\textbf{Resource Availability:}\\
The source code and artifact of this paper have been made publicly available at \url{\kddavailabilityurl}, with the GitHub repository at \url{https://github.com/happysnail06/SAGEO_Arena}.
\endgroup
}

\section{Introduction}
\label{sec:introduction}
\input{MainText/010_introduction}

\section{Related Work}
\label{sec:relatework}
\input{MainText/020_relatedwork}

\section{\proposed}
\label{sec:benchmark}
\input{MainText/030_benchmark}

\section{Experimental Setup}
\label{sec:experimental_setup}
\input{MainText/040_experimental_setup}

\section{Results \& Discussion}
\label{sec:results}
\input{MainText/050_results}

\section{Conclusion}
\label{sec:conclusion}
\input{MainText/060_conclusion}

\section*{Acknowledgments}
\label{sec:acknowledge}
\input{MainText/061_acknowledgements}

\bibliographystyle{ACM-Reference-Format}
\balance
\bibliography{bibliography}

\appendix
\input{MainText/070_appendix}

\end{document}

%% file: dfn.tex
\newcommand{\proposed}{\textsc{SAGEO Arena}\xspace}

%% file: MainText/000_abstract.tex
Search-Augmented Generative Engines (SAGE) have emerged as a new paradigm for information access, bridging web-scale retrieval with generative capabilities to deliver synthesized answers.
This shift has fundamentally reshaped how web content gains exposure online, giving rise to Search-Augmented Generative Engine Optimization (SAGEO), the practice of optimizing web documents to improve their visibility in AI-generated responses. 
Despite growing interest, no evaluation environment currently supports comprehensive investigation of SAGEO. 
Existing benchmarks lack end-to-end visibility evaluation of optimization strategies, operating on pre-determined candidate documents and abstracting away retrieval and reranking stages in SAGE.
They also discard structural information (e.g., schema markup) present in real web documents, overlooking the rich signals that search systems actively leverage in practice.
Motivated by these gaps, we introduce \proposed, a realistic and reproducible environment for stage-level SAGEO analysis.
\proposed integrates a full generative search pipeline over a large-scale corpus of web documents with rich structural information, enabling the first empirical analysis of how optimization signals propagate from retrieval to generation. 
Our findings reveal that existing approaches remain largely impractical, often degrading visibility in retrieval and reranking.
Leveraging structural information helps mitigate these limitations, yet effective SAGEO requires tailoring optimization to each pipeline stage.
Overall, our benchmark provides a practical foundation for advancing SAGEO.

%% file: MainText/010_introduction.tex
\input{Figures/010_intro_fig}
With the rapid advancement of Large Language Models (LLMs), Search-Augmented Generative Engines (SAGE)~\cite{chen2024benchmarking, wang2024searching,10.1145/3774904.3792724, kim2025bespoke} have emerged as a new paradigm for information access.
{By integrating large-scale retrieval with generative capabilities, these systems perform \textit{generative search}, providing synthesized answers that directly address user queries.}
As an increasing portion of web traffic now originates from such AI-generated answers~\cite{chen2026navigating}, how content gains exposure online is fundamentally changing.
This shift has given rise to Search-Augmented Generative Engine Optimization (SAGEO), the practice of optimizing web documents to improve their visibility (i.e., presence and contribution) in AI-generated responses.

SAGEO naturally extends the core objective of traditional Search Engine Optimization (SEO)~\cite{gudivada2015understanding}.
For decades, optimization has focused on improving retrieval and ranking on Search Engine Results Pages (SERPs), primarily through on-page factors such as titles and metadata~\cite{sharma2019brief}.
With the emergence of SAGE, optimization has increasingly shifted toward a generation-centric perspective, studying how a document is favored and cited in model responses~\cite{chen2025generative}.
A prominent line of study is Generative Engine Optimization (GEO)~\cite{aggarwal2024geo}, focusing on presentational modifications to improve visibility at the generation stage.
Despite this shift, generative search often builds on traditional search systems.
For example, MS Copilot relies on Bing Search to ground its responses~\cite{microsoft2026copilotsearch}, meaning that documents must first be surfaced by the underlying search pipeline before they can enter the LLM's context.
Thus, SEO and GEO must be jointly addressed to achieve successful optimization.
In light of this, we clearly distinguish SAGEO as optimization that targets the full generative search pipeline, from retrieval and reranking to generation (Figure ~\ref{fig:intro_fig}).
This motivates a systematic study of what constitutes effective SAGEO and how optimization signals propagate and interact across stages in generative search.

Despite growing interest, there exists no evaluation environment that enables such investigation.
Existing benchmarks~\cite{aggarwal2024geo, wu2025generative, puerto2025c, chen2025beyond} tend to oversimplify real web environments, leaving it unclear whether current optimization strategies are effective in practical settings.
Specifically, there are two major limitations: 
\textbf{(1) Lack of end-to-end evaluation.}
Current benchmarks predominantly operate on pre-determined {candidate documents,} abstracting away the retrieval and reranking stages that precede generation.
It remains unclear whether current optimization strategies trade off or benefit performance at earlier search stages, making stage-level analysis of SAGEO essential.
\textbf{(2) Loss of structural information.}
Existing benchmarks operate solely on plain body text, discarding structural information present in real web documents (e.g., schema markup).
In practice, real search systems do not determine visibility from body text alone. 
Structural information continues to play an important role in search-stage visibility, while similar structured representations, such as markdown summaries, have also been shown to improve generation-stage visibility~\cite{puerto2025c}.
Yet no existing benchmark preserves structural information, leaving its impact across the generative search pipeline largely unexplored.

Motivated by these gaps, we introduce \proposed, a realistic and reproducible benchmark designed for stage-level SAGEO analysis.
Crucially, our benchmark is distinguished in two ways:
\textbf{(1) {Comprehensive generative search environment.}}
Unlike existing benchmarks that assume fixed retrieval context, \proposed~integrates a full generative search pipeline where documents are dynamically retrieved, ranked, and passed to generation.
Each stage is modular and configurable, allowing researchers to trace how optimization signals propagate across the complete search process.
\textbf{(2) Extensive corpus with rich structural annotations.}
We curate a corpus of 170k web documents spanning 9 domains.
Following guidelines published by commercial search engines, we extract both body text and rich structural information that SAGEO must navigate in practice (e.g., meta descriptions, headings, and schema markup).
Overall, we aim to construct an environment that closely reflects real-world deployment, paving the way for optimization strategies that generalize beyond controlled settings.
\input{Tables/020_benchmark_comparison_table}

We conduct extensive experiments on \proposed~to evaluate how optimization strategies affect document visibility across the full generative search pipeline.
Our analysis reveals that body text optimization alone, the predominant focus of prior work, remains largely insufficient in realistic generative search settings.
It provides only marginal gains in generation-stage visibility and, more importantly, degrades retrieval performance, causing optimized documents to drop out of the retrieval results.
By contrast, structural information helps mitigate this degradation by preserving signals that support document visibility in early pipeline stages.
We also find that the two scopes play fundamentally complementary roles, with structural information driving retrieval and informative body text remaining a key factor for reranking and generation.
Through in-depth analysis across domains, citation behavior, and optimization models, we further show that each pipeline stage prioritizes different document qualities, motivating the need for stage-aware optimization.
Guided by these findings, we propose stage-aware SAGEO, a method that tailors optimization strategies to each stage, achieving the strongest overall visibility gain among all evaluated strategies.
To summarize, our contributions are as follows:
\setlength{\leftmargini}{15pt}
\setlength{\itemsep}{5pt}
\begin{itemize}
    \item We introduce \proposed, the first benchmark that enables stage-level visibility evaluation of SAGEO. By integrating a full generative search pipeline with a large-scale corpus preserving rich structural information, \proposed~bridges the gap between existing benchmarks and real-world SAGE.
    
    \item We provide the first empirical analysis of structural information optimization, studying how structural signals that generative search systems must navigate in practice influence visibility.

    \item Through extensive experiments, we show that current optimization approaches are insufficient in realistic settings. To address this, we introduce stage-aware SAGEO, providing actionable guidance to further facilitate SAGEO research.
\end{itemize}

%% file: Figures/010_intro_fig.tex
\begin{figure}[!t]
\centering
\includegraphics[width=1.0\linewidth]{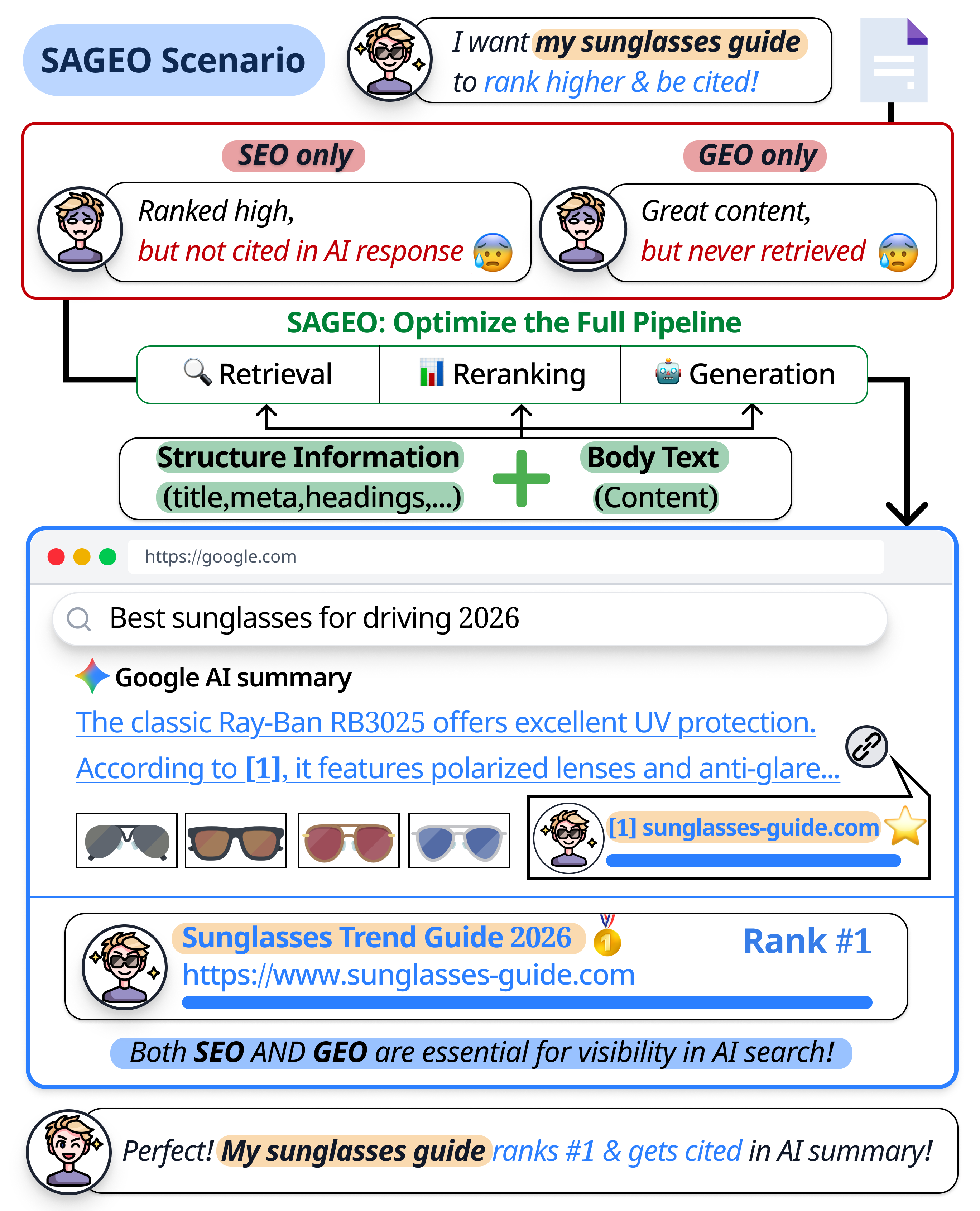}
\caption{
Optimizing for ranking alone (SEO) or citation alone (GEO) causes failure at either retrieval/reranking or generation. SAGEO jointly optimizes structural information and body text across the full pipeline to succeed at both stages.
}
\label{fig:intro_fig}
\end{figure}

%% file: Tables/020_benchmark_comparison_table.tex
\definecolor{darkgreen}{HTML}{006400}
\begin{table*}[ht]
\caption{A comparison of \proposed to existing benchmarks.}
\centering
\label{tab:benchmark_comparison_table}
\resizebox{\textwidth}{!}{%
\begin{tabular}{lcccccccccc}
\toprule
\multirow{2.5}{*}{\textbf{Benchmark}} & \multicolumn{4}{c}{\textbf{Environment}} & \multicolumn{2}{c}{\textbf{Document Details}} & \multicolumn{3}{c}{\textbf{Evaluation}} \\
\cmidrule(lr){2-5} \cmidrule(lr){6-7} \cmidrule(lr){8-10}
 & \textbf{Doc. Corpus} & \textbf{Retrieval} & \textbf{Reranking} & \textbf{Generation} & \textbf{Structure Info.} & \textbf{Body Text} & \textbf{Search} & \textbf{Generation} & \textbf{Visibility Metric}\\
\midrule
GEO-Bench \cite{aggarwal2024geo} & - & \color{red}\ding{55} & \color{red}\ding{55} & \color{darkgreen}\ding{51} & \color{red}\ding{55} & \color{darkgreen}\ding{51} & \color{red}\ding{55} & \color{darkgreen}\ding{51} & Word Count \\
AutoGEO \cite{wu2025generative} & - & \color{red}\ding{55} & \color{red}\ding{55} & \color{darkgreen}\ding{51} & \color{red}\ding{55} & \color{darkgreen}\ding{51} & \color{red}\ding{55} & \color{darkgreen}\ding{51} & Word Count, Utility \\
C-SEO Bench \cite{puerto2025c} & - & \color{red}\ding{55} & \color{red}\ding{55} & \color{darkgreen}\ding{51} & \color{red}\ding{55} & \color{darkgreen}\ding{51} & \color{red}\ding{55} & \color{darkgreen}\ding{51} & Citation Rank \\
CC-GSEO-Bench \cite{chen2025beyond} & - & \color{red}\ding{55} & \color{red}\ding{55} & \color{darkgreen}\ding{51} & \color{red}\ding{55} & \color{darkgreen}\ding{51} & \color{red}\ding{55} & \color{darkgreen}\ding{51} & Influence \\
\midrule
\textbf{\proposed (Ours)} & 170K & \color{darkgreen}\ding{51} & \color{darkgreen}\ding{51} & \color{darkgreen}\ding{51} & \color{darkgreen}\ding{51} & \color{darkgreen}\ding{51} & \color{darkgreen}\ding{51} & \color{darkgreen}\ding{51} & Hit Rate, Rank Change \\
\bottomrule
\end{tabular}%
}
\end{table*}

%% file: MainText/020_relatedwork.tex
\subsection{Generative Search Engine Optimization}
With the rise of LLMs, generative search~\cite{zheng2025deepresearcher,li2026webthinker,jin2025search,chen2025research} has become a cornerstone of modern information access, offering synthesized answers grounded in retrieved documents.
This shift has introduced new visibility challenges for content creators, motivating research on optimization strategies tailored to generative engines.
Existing work can be broadly categorized into two directions:
(1) \textit{Black-hat} approaches explore adversarial methods that exploit model behavior through malicious content injection~\cite{nestaas2025adversarial,kumar2024manipulating,pfrommer2024ranking}.
In contrast, (2) \textit{white-hat} approaches focus on cooperative content modification, formalized as GEO ~\cite{aggarwal2024geo}.
Building on this foundation, subsequent work~\cite{wu2025generative, puerto2025c, chen2025beyond} has advanced GEO research by proposing diverse optimization strategies and evaluation dimensions (Table ~\ref{tab:benchmark_comparison_table}).
Nevertheless, existing evaluation protocols often rely on oversimplified environments, predominantly assessing optimization effects at the generation stage with fixed document candidates.
In other words, they assume that the target document remains within the LLM context after optimization, without testing whether the optimized document can still be retrieved and reranked into that context. 
Thus, how optimization signals propagate through the full generative pipeline remains largely unexplored.
Motivated by these, this study provides a comprehensive evaluation environment that explicitly models the end-to-end generative search pipeline.

\subsection{Comparison of SEO and GEO}
Traditional SEO has long served as the dominant framework for online visibility~\cite{raifer2017information, sharma2019brief, goren2020ranking, kurland2022competitive}, aiming to improve a source's position on SERP.
Following established guidelines from leading search engine providers~\cite{bing2025webmastersguidelines, google_how_search_works}, SEO has long emphasized structural information that helps pages be interpreted and surfaced, including titles, meta descriptions, headings, and schema markup.
In contrast, GEO research predominantly focuses on modifying body text in the generation stage~\cite{chen2025role}, investigating rewriting strategies that adjust textual properties such as fluency, technicality, and ease of understanding.
However, recent empirical findings~\cite{puerto2025c} challenge this generation-centric perspective, demonstrating that a document's position determined in the retrieval and reranking stage plays a far more dominant role in determining its visibility in the final response.
Notably, document position is often shaped by structural information emphasized in traditional SEO, which raises a fundamental question: \textit{do these signals also contribute to visibility in generative engines?}
Existing benchmarks offer limited insight into this question, 
and we bridge this gap by jointly studying how SEO and GEO signals interact across the full generative search pipeline.

\subsection{Structured Web Information Understanding}
The World Wide Web is fundamentally built upon HyperText Markup Language (HTML). 
Beyond body content, HTML documents encode structural information through elements such as title tags, meta descriptions, heading hierarchies, and schema markup.
These structural attributes have long influenced how search engines crawl, index, and rank documents~\cite{10.1145/3447535.3462479}.
Recent studies~\cite{Deng2022DOMLMLG, Guo2022WebformerPW, Wang2022WebFormerTW, Tan2024HtmlRAGHI} demonstrate that structural information helps LLMs better understand web content, showing that structural representations in HTML complement plain text in Retrieval-Augmented Generation (RAG) systems.
Moreover, there is evidence indicating that commercial SAGEs, such as MS Copilot and Google Search, similarly leverage structural information when interpreting and surfacing web documents~\cite{kumar2025ai, bing2025webmastersguidelines, google2025seostarterguide}.
These findings motivate incorporating structural information into SAGEO research, but no existing benchmark systematically evaluates its impact.
We address this gap by proposing a novel benchmark that preserves the structural elements in web documents, enabling more realistic evaluation of optimization strategies in real-world generative search settings.

%% file: MainText/030_benchmark.tex
\input{Figures/030_main_fig}

\noindent\textbf{Problem Setting.}
We consider a search-augmented generative engine that answers user queries by retrieving relevant {web documents} and synthesizing a response grounded in the retrieved content.
{The pipeline follows the well-established RAG paradigm~\cite{lewis2020retrieval}}, consisting of a retriever, reranker, and generator. 
In this setting, a document must first be retrieved and ranked before it can influence the generated response. 
This creates an inherent dependency between traditional SEO concerns (retrieval and ranking) and emerging GEO concerns (generation). 
Our benchmark is designed to capture this full pipeline, enabling thorough evaluation of optimization strategies across all three stages.

\medskip
\noindent\textbf{Design Principles.}
Our core challenge lies in reproducing an environment that reflects real-world search-augmented generative engines.
To achieve this, we {(1)} adopt a standardized search-augmented generation pipeline following established practices~\cite{gao2023retrieval, rau2024bergen, googlecloud_vertexai_rag_overview_2026} and {(2)} preserve structural information explicitly recommended by commercial search engines for optimization~\cite{google_how_search_works,bing2025webmastersguidelines,bing2025howbingdelivers}.

\medskip
\noindent\textbf{Task Formulation.}
Let $q$ denote a user query and $\mathcal{D}$ a document corpus. Given $q$, the system produces a response $A_q$ through three stages:
(1) a retriever $\mathcal{R}$ returns a candidate set $\mathcal{D}_q = \mathcal{R}_k(q, \mathcal{D})$ containing the top-$k$ documents; 
(2) a reranker $\mathcal{F}$ reorders $\mathcal{D}_q$ by relevance, producing a ranked list $\mathcal{D}_q^* = \mathcal{F}(q, \mathcal{D}_q)$;
(3) a generator $\mathcal{G}$ produces a response $A_q = \mathcal{G}(q, \mathcal{D}_q^*)$ with inline citations.
\begin{equation}
A_q \;=\; \mathcal{G}\!\Big(q,\; \mathcal{F}\big(q,\; \mathcal{R}_k(q,\mathcal{D})\big)\Big).
\end{equation}

\noindent
The visibility of each document $d_i \in \mathcal{D}_q^*$ is reflected by whether and to what extent it is cited in $A_q$.
Let $d^{\mathrm{tgt}} \in \mathcal{D}$ denote a target document relevant to the query $q$.
SAGEO aims to optimize $d^{\mathrm{tgt}}$, without prior knowledge of the query $q$, such that it (1) is retrieved into $\mathcal{D}_q$, (2) ranks highly in $\mathcal{D}_q^*$, and (3) maximizes visibility in $A_q$.

\subsection{Corpus Construction}
To create a large-scale corpus with strong domain diversity, we curate queries from nine established information retrieval datasets spanning a broad range of domains (Table~\ref{tab:benchmark_stats}). 
We sample 300 queries from each dataset, yielding 2,700 unique queries in total.
Following prior approaches, we retrieve up to 100 search results per query using the Google Custom Search API.
We then crawl each URL to extract both body text and rich structural information (Section~\ref{sec:structural_fields}).
After filtering documents with malformed content, the final corpus contains 171,003 unique web documents.
On average, each query is associated with 63 candidate documents.
{The 300 sampled queries per domain serve as test queries for evaluating optimization.}
Detailed statistics of \proposed are provided in Appendix~\ref{appendix:dataset}.
%

\subsection{Structural Information Extraction}
\label{sec:structural_fields}
Unlike prior benchmarks that extract only body text, we explicitly preserve structural information embedded in real-world web documents.
However, such documents exhibit substantial diversity and noise~\cite{10.1145/956750.956785, Uzun2013AHA}, making it impractical to preserve all structural information for systematic evaluation. 
We therefore selectively retain structural fields consistently emphasized in public search engine guidelines~\cite{bing2025webmastersguidelines}, enabling a realistic yet controlled study of how these elements influence search-augmented generative engines.
\setlength{\leftmargini}{10pt}
\setlength{\itemsep}{2pt}
\begin{itemize}
    \item \textbf{Title.} The document title provides a concise topical summary of the page. It serves as a primary signal of the page’s main topic and intent, playing a key role in initial relevance assessment.
    
    \item \textbf{Meta Description.} A concise page summary intended for search result snippets. This field offers a compact yet informative description of page content that complements the title.

    \item \textbf{Headings.} Hierarchical section headers (H1--H6) that signal topical structure within a document. Search engines treat headings as strong indicators of what each section is about, often weighting them higher than surrounding body text.
    
    \item \textbf{Schema/JSON-LD.}  Structured data markup that explicitly defines entities, attributes, and relationships in a machine-readable format. Search engines rely on this markup to interpret page semantics that cannot be inferred from plain text alone.
    
    \item \textbf{Body Text.} The main textual content of the document, representing the primary source of information.
    
\end{itemize}

\noindent
Notably, merging these fields into a unified text representation at the indexing stage would obscure the distinct signals carried by each component. Therefore, search-augmented generative engines broadly index these fields as separate components and combine them later during relevance scoring~\cite{Robertson2004SimpleBE}. We adopt the same practice, preserving field-level separation to enable fine-grained analysis of how each component contributes to visibility across the pipeline.

\subsection{Search-Augmented Generative Engine}
\label{sec:pipeline}
We implement a full generative search engine pipeline with retrieval, reranking, and generation.
Each stage reflects real-world practices while remaining modular for controlled experimentation.

\smallskip
\noindent\textbf{{Document Representation}.}
In our setting, a web document consists of two types of content: structural information and body text. The structural information of a document, denoted as $\mathcal{S}(d) = \{s_1, \ldots, s_m\}$, is a set of structural fields corresponding to the elements described in Section~\ref{sec:structural_fields} (e.g., title, meta description). The body text of a document, denoted as $\mathcal{P}(d) = \{p_1, \ldots, p_n\}$, is a set of passages chunked from the body text following standard practice. A document can therefore be represented as $(\mathcal{S}(d), \mathcal{P}(d))$.

\smallskip
\noindent\textbf{{Semantic Unit}.}
In generative search systems, documents are often chunked and retrieved at the passage level. However, structural information of a web document is shared across all passages, making it essential to bridge this gap between document-level and passage-level information for effective retrieval. We therefore pair each passage $p_i \in \mathcal{P}(d)$ with the associated structural information $\mathcal{S}(d)$, creating a semantic unit $\mathcal{U}(p_i) = \mathcal{S}(d) \cup \{p_i\}$ that serves as a passage representation of $p_i$ during retrieval and reranking.

\smallskip
\noindent\textbf{{Retriever.}}
We employ BM25~\cite{10.1561/1500000019}, a widely adopted lexical retrieval method, as our retriever $\mathcal{R}$.
This choice reflects the practical role of lexical retrieval as an efficient and scalable first-stage candidate generator in web-scale search systems.
For each semantic unit $\mathcal{U}(p)$, we build a separate BM25 index for every element $u \in \mathcal{U}(p)$.
Given a query $q$, we score $q$ against each $u$ independently, and aggregate the resulting ranks via reciprocal rank fusion to obtain the retrieval score of passage $p$:
\begin{equation}
\text{score}_{\mathcal{R}}(q, p) = \sum_{u \,\in\, \mathcal{U}(p)} \frac{1}{\kappa + \text{rank}(q, u)},
\end{equation}
\noindent where $\text{rank}(q, u)$ is the rank of element $u$, and $\kappa$ is a constant.
The top-$k$ passages are selected as candidates for reranking.

\smallskip
\noindent\textbf{Reranker.}
The retrieved candidates are reranked using a cross-encoder that scores each query-passage pair $(q, p)$ independently.
Following practices in production systems (e.g., Vespa AI\footnote{\url{https://vespa.ai/}}, Google Vertex AI\footnote{\url{https://cloud.google.com/vertex-ai}}), we score relevance at the passage level rather than over entire documents. 
This approach handles cases where only specific sections of a long document are relevant to the query.

\smallskip
\noindent\textbf{Generator.}
The top reranked candidates are provided to the generation model along with their structural fields.
The model is prompted to cite sources, enabling measurement of document-level visibility.

\smallskip
\noindent By grounding our design in a complete pipeline with structurally enriched documents, \proposed enables analyses beyond the scope of existing benchmarks. 
Specifically, it allows (1) stage-wise evaluation of diverse strategies, (2) component-level ablations, and (3) testing robustness and generalization across realistic settings.

\subsection{Stage-level Visibility Evaluation}
We describe how \proposed measures document visibility across each pipeline stage and quantifies the effect of optimization.

\medskip
\noindent\textbf{Evaluation Pipeline.}
\label{sec:evaluation_pipeline}
As shown in Figure~\ref{fig:main_fig}, we establish a baseline as follows.
For each test query $q$, we first execute the full generative search pipeline in \proposed.
Among the documents that reach the generation stage (i.e., ranked within the top-$k$ at the reranking stage), we randomly select one as the target document $d^{\mathrm{tgt}}$, yielding the baseline instance $(q, d^{\mathrm{tgt}})$.
The target document $d^{\mathrm{tgt}}$ is then optimized using a specified strategy and reindexed into the corpus following the process described in Section~\ref{sec:pipeline}.
We then re-execute the search pipeline with the same query and compare the document's visibility against the baseline at each stage.
This enables evaluating optimization effects under realistic, stage-level settings, rather than within a fixed retrieval context like prior benchmarks.

\medskip
\noindent\textbf{{Tracking Target Document.}}
Since search-augmented generative engines internally operate at the passage level,  multiple passages from the same document may appear in the candidate list as they are scored independently.
This makes tracking the visibility of a target document an inherent challenge.
To address this, we define the rank of a target document $d^{\mathrm{tgt}}_q$ as the highest rank achieved by any of its associated passages at each pipeline stage:
\begin{equation}
\label{eq:doc_rank}
\mathrm{rank}(d^{\mathrm{tgt}}_q) = \min_{p \in \mathcal{P}(d^{\mathrm{tgt}}_q)} \mathrm{rank}(p),
\end{equation}
where $\mathcal{P}(d^{\mathrm{tgt}}_q)$ denotes the set of passages derived from $d^{\mathrm{tgt}}_q$.
The passage with the highest rank serves as the visibility indicator of the associated document at each stage, and we measure optimization effectiveness in two ways: (1) whether the target document appears within the top-$k$ candidates at each stage, and (2) how the rank of the target document shifts before and after optimization.

\medskip
\noindent\textbf{Metrics.}
For each test query $q \in Q$, we execute the pipeline and track the
rank of its associated target document $d^{\mathrm{tgt}}_q$ via
Equation~\eqref{eq:doc_rank}.
We then evaluate visibility at each stage using two complementary metrics:
{(1)} \textsc{Hit Rate}\xspace (H@$k$), measuring the proportion of queries for which the target document appears within the top-$k$ candidates.
For retrieval and reranking, it is defined by
\begin{equation}
\mathrm{H@}k = \frac{1}{|Q|} \sum_{q \in Q} \mathbb{I}\left(\mathrm{rank}(d^{\mathrm{tgt}}_q) \leq k\right),
\end{equation}
where $\mathbb{I}(\cdot)$ returns 1 if the condition is satisfied and 0 otherwise.
For generation, we instead use \textsc{Citation Rate}\xspace, the
proportion of queries where the target document is cited in the final
response.
{(2)} \textsc{Rank Change}\xspace, the average positional shift of the target document between baseline and optimized settings:
\begin{equation}
\Delta \mathrm{Rank} = \frac{1}{|Q|} \sum_{q \in Q} \left( \mathrm{rank}_{\mathrm{base}}(d^{\mathrm{tgt}}_q) - \mathrm{rank}_{\mathrm{SAGEO}}(d^{\mathrm{tgt}}_q) \right),
\end{equation}
\noindent
where $\mathrm{rank}_{\mathrm{base}}$ and $\mathrm{rank}_{\mathrm{SAGEO}}$ represent the target document rank before and after optimization, respectively.
{Documents not appearing in the top-$k$ candidates either at the retrieval or reranking stage are assigned a default rank of $k+1$.
We set $k=100$ for retrieval and $k=10$ for reranking. 
For generation, rank at this stage is defined as the citation order in which a source first appears in the response.

%% file: Figures/030_main_fig.tex
\begin{figure*}[t]
    \centering
    \includegraphics[width=0.89\textwidth]{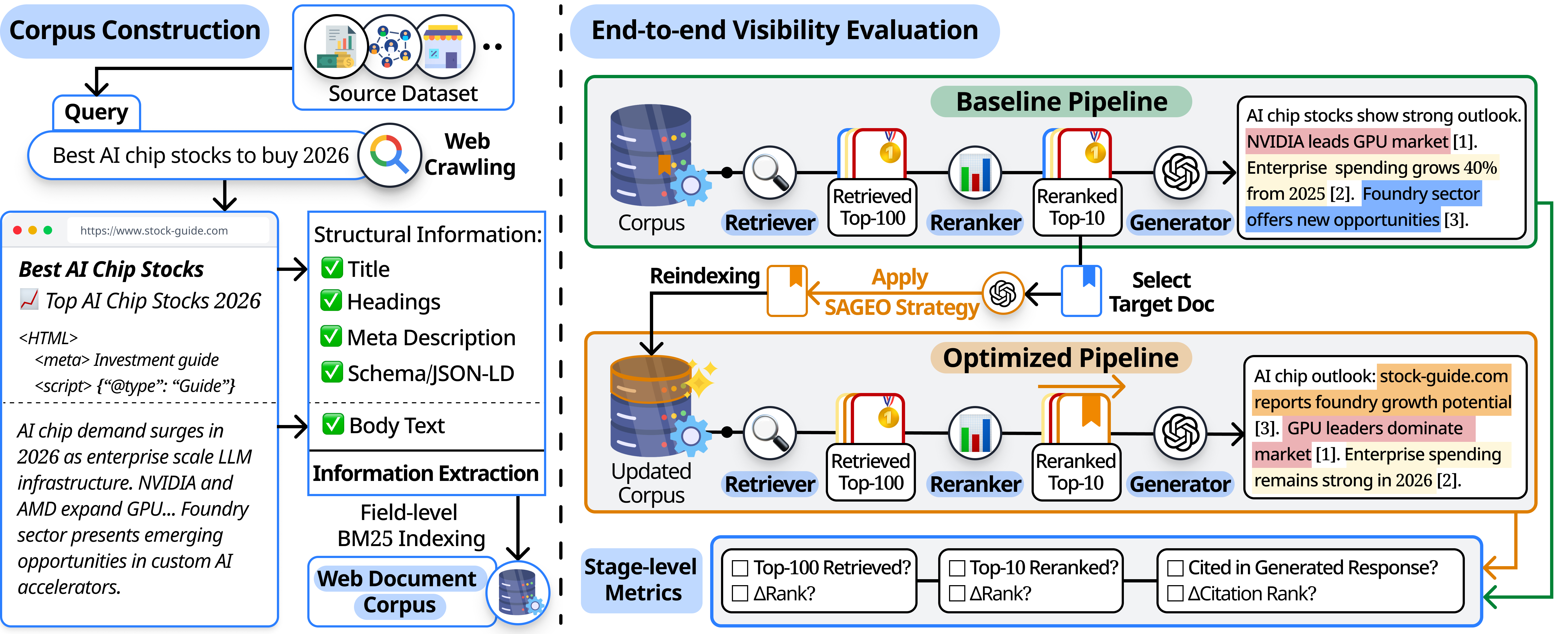}
    \vspace{-0.2cm} 
    \caption{Overview of \proposed. During document corpus construction, we extract structural information (title, meta description, headings, schema) and body text from web documents. During SAGEO evaluation, we first execute the pipeline with each test query and randomly select a target document from those that reach the generation stage (top-$k$ after reranking) as the baseline. The target document is then optimized, re-indexed into the corpus, and the pipeline is re-executed with the same query, tracking whether the document maintains, gains, or loses visibility at each stage (retrieval, reranking, generation).}
    \label{fig:main_fig}
    \vspace{-0.3cm}
\end{figure*}

%% file: MainText/040_experimental_setup.tex
\input{Tables/050_main_table}
\noindent\textbf{{Optimization Strategies.}}
We evaluate 10 {LLM-based optimization} strategies derived from prior research.
Eight are adapted from~\cite{aggarwal2024geo}, each targeting a specific stylistic or content modification to improve document visibility.
These include strategies that adjust tone and readability (Authoritative, Fluency, Easy Language), add supporting evidence (Cite Sources, Quotation, Statistics), and diversify vocabulary (Technical Terms, Unique Words).
We additionally include All-in-One~\cite{puerto2025c}, which applies all eight strategies simultaneously along with structural formatting such as bolding and improved layout, and AutoGEO~\cite{wu2025generative}, which leverages preference rules learned from generative engine behavior to guide document optimization.

\begin{itemize}
\item \textbf{Authoritative:} Modifies text style to be more persuasive, making statements definitive while maintaining factual accuracy.

\item \textbf{Cite Sources:} Adds inline citations to external sources, inserting brief references to enhance reliability and trustworthiness.

\item \textbf{Fluency:} Improves grammatical correctness by refining sentence structure and transitions without altering content meaning.

\item \textbf{Quotation:} Incorporates quotations from authoritative figures or sources with proper attribution to provide external validation.

\item \textbf{Easy Language:} Simplifies by replacing complex vocabulary with accessible alternatives while preserving information.

\item \textbf{Statistics:} Adds quantitative data and numerical facts inline within sentences to make claims more concrete and credible.

\item \textbf{Technical Terms:} Introduces domain-specific terminology to present content in a more expert and authoritative manner.

\item \textbf{Unique Words:} Enriches vocabulary by incorporating less common words to signal higher content quality and specialization.

\item \textbf{All-in-One:} Combines all eight base strategies, then enhances text structure by bolding key features and improving layout to increase information accessibility and visual attractiveness.

\item \textbf{AutoGEO:} Applies optimization rules learned from analyzing generative engine citation patterns to guide document rewriting.

\end{itemize}

\medskip
\noindent\textbf{{Pipeline Configuration.}}
We use BM25 as the retriever, Qwen3-Reranker-4B~\cite{qwen3embedding} as the reranker, and GPT-5-mini as the generator in the default SAGEO Arena pipeline.
Further implementation details of the pipeline and optimization are described in Appendix ~\ref{appendix:implementation_details}.

%% file: Tables/050_main_table.tex
\definecolor{gain-strong}{RGB}{197,224,246}      
\definecolor{gain-moderate}{RGB}{214,234,248}    
\definecolor{gain-mild}{RGB}{232,242,251}        
\definecolor{neutral}{RGB}{245,245,245}          
\definecolor{drop-mild}{RGB}{254,242,242}        
\definecolor{drop-moderate}{RGB}{252,229,229}    
\definecolor{drop-severe}{RGB}{248,205,205}      

\definecolor{pct-pos-strong}{RGB}{31,119,180}    
\definecolor{pct-pos-moderate}{RGB}{100,150,200} 
\definecolor{pct-pos-mild}{RGB}{150,180,220}     
\definecolor{pct-neg-mild}{RGB}{245,170,170}     
\definecolor{pct-neg-moderate}{RGB}{235,130,130} 
\definecolor{pct-neg-strong}{RGB}{224,79,80}     
\definecolor{pct-zero}{RGB}{150,150,150}         

\newcommand{\gs}[1]{\cellcolor{gain-strong}{#1}}
\newcommand{\gm}[1]{\cellcolor{gain-moderate}{#1}}
\newcommand{\gl}[1]{\cellcolor{gain-mild}{#1}}
\newcommand{\nn}[1]{\cellcolor{neutral}{#1}}
\newcommand{\dm}[1]{\cellcolor{drop-mild}{#1}}
\newcommand{\dd}[1]{\cellcolor{drop-moderate}{#1}}
\newcommand{\ds}[1]{\cellcolor{drop-severe}{#1}}

\newcommand{\base}{\textcolor{gray}{\scriptsize -\,-\,\%}}

\newcommand{\pps}[1]{\makebox[2.5em][l]{\textcolor{pct-pos-strong}{\small #1}}}
\newcommand{\ppm}[1]{\makebox[2.5em][l]{\textcolor{pct-pos-moderate}{\small #1}}}
\newcommand{\ppl}[1]{\makebox[2.5em][l]{\textcolor{pct-pos-mild}{\small #1}}}
\newcommand{\nps}[1]{\makebox[2.5em][l]{\textcolor{pct-neg-strong}{\small #1}}}
\newcommand{\npm}[1]{\makebox[2.5em][l]{\textcolor{pct-neg-moderate}{\small #1}}}
\newcommand{\npl}[1]{\makebox[2.5em][l]{\textcolor{pct-neg-mild}{\small #1}}}
\newcommand{\zp}[1]{\makebox[2.5em][l]{\textcolor{pct-zero}{\small #1}}}
\newcommand{\bp}{\makebox[2.5em][l]{\textcolor{pct-zero}{\small -\,-\,\%}}}

\begin{table*}[ht]
\caption{Effectiveness of SAGEO strategies. We report Hit Rate and $\Delta$Rank for each optimization target at each stage. Percentage values indicate relative change from the pre-optimization baseline. Positive $\Delta$Rank indicates rank improvement.
H@10 at reranking is 1.00 in the baseline, as all target documents are selected {from the top-10 candidates at the reranking stage.}}
\vspace{-0.2cm} 
\centering
\label{tab:main_results}
\setlength{\tabcolsep}{1pt}
\renewcommand{\arraystretch}{1.15}
\footnotesize
\resizebox{\textwidth}{!}{%
\begin{tabular}{l cc cc cc | cc cc cc | cc cc cc}
\toprule
\multirow{3}{*}[-0.5em]{\textbf{Strategy}} 
& \multicolumn{6}{c|}{\textbf{Body Text only}} 
& \multicolumn{6}{c|}{\textbf{Structural Information only}} 
& \multicolumn{6}{c}{\textbf{Both}} \\
\cmidrule(lr){2-7} \cmidrule(lr){8-13} \cmidrule(lr){14-19}
& \multicolumn{2}{c}{Retrieval} & \multicolumn{2}{c}{Reranking} & \multicolumn{2}{c|}{Generation} 
& \multicolumn{2}{c}{Retrieval} & \multicolumn{2}{c}{Reranking} & \multicolumn{2}{c|}{Generation} 
& \multicolumn{2}{c}{Retrieval} & \multicolumn{2}{c}{Reranking} & \multicolumn{2}{c}{Generation} \\
\cmidrule(lr){2-3} \cmidrule(lr){4-5} \cmidrule(lr){6-7} 
\cmidrule(lr){8-9} \cmidrule(lr){10-11} \cmidrule(lr){12-13} 
\cmidrule(lr){14-15} \cmidrule(lr){16-17} \cmidrule(lr){18-19}
& H@20 & $\Delta$Rank & H@10 & $\Delta$Rank & Cite & $\Delta$Rank 
& H@20 & $\Delta$Rank & H@10 & $\Delta$Rank & Cite & $\Delta$Rank 
& H@20 & $\Delta$Rank & H@10 & $\Delta$Rank & Cite & $\Delta$Rank \\
\midrule
\rowcolor{gray!5}
\textit{Baseline} & 0.58 \bp & \nn{-} & 1.00 \bp & \nn{-} & 0.50 \bp & \nn{-} & 0.58 \bp & \nn{-} & 1.00 \bp & \nn{-} & 0.50 \bp & \nn{-} & 0.58 \bp & \nn{-} & 1.00 \bp & \nn{-} & 0.50 \bp & \nn{-} \\
\midrule
Auth.     & \textbf{0.57} \npl{-1\%} & \dm{\textbf{-0.20}} & \textbf{0.91} \npl{-9\%} & \dm{-0.24} & 0.49 \npl{-3\%} & \nn{-0.10} & 0.68 \ppl{+18\%} & \gl{+1.60} & 0.81 \nps{-19\%} & \dm{-0.59} & 0.49 \npl{-3\%} & \nn{+0.01} & 0.69 \ppl{+19\%} & \gl{+0.47} & 0.77 \nps{-23\%} & \dm{-0.80} & 0.48 \npl{-3\%} & \nn{+0.01} \\
Cite      & 0.56 \npl{-3\%} & \dm{-1.50} & 0.87 \npm{-13\%} & \dm{-0.45} & 0.48 \npl{-4\%} & \dm{-0.13} & 0.74 \ppm{+28\%} & \gs{+6.05} & 0.86 \npm{-14\%} & \dm{-0.17} & 0.52 \ppl{+2\%} & \gl{+0.25} & 0.66 \ppl{+13\%} & \dm{-1.81} & 0.73 \nps{-26\%} & \dm{-0.99} & 0.47 \npl{-6\%} & \nn{-0.04} \\
EasyLang  & 0.52 \npl{-10\%} & \dd{-4.18} & 0.84 \npm{-16\%} & \dm{-0.54} & 0.49 \npl{-4\%} & \nn{-0.09} & 0.74 \ppm{+27\%} & \gs{+5.15} & 0.86 \npm{-14\%} & \gl{\textbf{+0.13}} & 0.53 \ppl{+5\%} & \gl{+0.38} & 0.71 \ppm{+22\%} & \gm{+3.76} & 0.80 \npm{-20\%} & \dm{-0.30} & 0.51 \ppl{+2\%} & \gl{+0.34} \\
Fluency   & \textbf{0.57} \npl{-1\%} & \dm{-0.71} & \textbf{0.91} \npl{-9\%} & \dm{\textbf{-0.18}} & \textbf{0.50} \npl{-1\%} & \nn{\textbf{-0.01}} & \textbf{0.75} \ppm{+30\%} & \gs{\textbf{+6.62}} & \textbf{0.88} \npm{-12\%} & \nn{+0.08} & 0.53 \ppl{+5\%} & \gl{+0.37} & 0.72 \ppm{+24\%} & \gm{+2.37} & 0.81 \npm{-19\%} & \dm{-0.42} & 0.49 \npl{-2\%} & \gl{+0.11} \\
Quote     & \textbf{0.57} \npl{-1\%} & \dm{-0.33} & 0.90 \npl{-10\%} & \dm{-0.36} & 0.47 \npl{-7\%} & \dm{-0.26} & 0.74 \ppm{+27\%} & \gs{+5.47} & 0.85 \npm{-15\%} & \dm{-0.28} & 0.53 \ppl{+4\%} & \gl{+0.30} & \textbf{0.79} \ppm{+35\%} & \gs{\textbf{+8.87}} & \textbf{0.85} \npm{-15\%} & \dm{\textbf{-0.29}} & 0.51 \ppl{+2\%} & \gl{+0.24} \\
Stats     & \textbf{0.57} \npl{-2\%} & \dm{-0.51} & 0.90 \npl{-10\%} & \dm{-0.33} & 0.48 \npl{-4\%} & \dm{-0.18} & \textbf{0.75} \ppm{+29\%} & \gs{+6.03} & 0.86 \npm{-14\%} & \dm{-0.15} & \textbf{0.54} \ppl{+7\%} & \gl{\textbf{+0.47}} & 0.71 \ppm{+22\%} & \gl{+1.53} & 0.80 \npm{-20\%} & \dm{-0.75} & 0.48 \npl{-5\%} & \nn{-0.08} \\
Tech.     & 0.50 \npm{-14\%} & \ds{-6.23} & 0.80 \npm{-20\%} & \dm{-1.03} & 0.47 \npl{-6\%} & \dm{-0.15} & 0.66 \ppl{+13\%} & \dm{-0.59} & 0.79 \nps{-21\%} & \dm{-0.61} & 0.51 \ppl{+1\%} & \gl{+0.22} & 0.62 \ppl{+6\%} & \dd{-2.39} & 0.64 \nps{-36\%} & \dd{-2.02} & 0.43 \npm{-14\%} & \dm{-0.44} \\
Unique    & 0.53 \npl{-8\%} & \dd{-3.47} & 0.86 \npm{-14\%} & \dm{-0.70} & 0.46 \npl{-8\%} & \dm{-0.33} & 0.69 \ppl{+19\%} & \gl{+1.09} & 0.81 \npm{-19\%} & \dm{-0.62} & 0.49 \npl{-4\%} & \nn{-0.06} & 0.66 \ppl{+13\%} & \dm{-0.16} & 0.75 \nps{-25\%} & \dm{-1.24} & 0.43 \npm{-13\%} & \dm{-0.48} \\
All-in-One  & 0.50 \npm{-14\%} & \ds{-5.93} & 0.83 \npm{-17\%} & \dm{-0.68} & 0.49 \npl{-2\%} & \nn{-0.03} & 0.71 \ppm{+22\%} & \gm{+2.44} & 0.82 \npm{-18\%} & \dm{-0.42} & 0.50 \npl{-1\%} & \gl{+0.12} & 0.69 \ppl{+17\%} & \gl{+1.96} & 0.77 \nps{-23\%} & \dm{-0.62} & \textbf{0.52} \ppl{+3\%} & \gl{\textbf{+0.38}} \\
AutoGEO   & 0.37 \nps{-36\%} & \ds{-22.35} & 0.58 \nps{-42\%} & \dd{-2.28} & 0.39 \nps{-22\%} & \dm{-0.78} & 0.60 \ppl{+4\%} & \ds{-6.68} & 0.73 \nps{-27\%} & \dm{-0.72} & 0.51 \zp{+0\%} & \gl{+0.31} & 0.44 \nps{-25\%} & \ds{-20.15} & 0.58 \nps{-42\%} & \dm{-1.93} & 0.44 \npm{-12\%} & \dm{-0.13} \\
\midrule
\textbf{Avg.} & 0.53 \npl{-9\%} & \dd{-4.54} & 0.84 \npm{-16\%} & \dm{-0.68} & 0.47 \npl{-6\%} & \dm{-0.21} & 0.71 \ppm{+22\%} & \gm{+2.72} & 0.83 \npm{-17\%} & \dm{-0.34} & 0.52 \ppl{+2\%} & \gl{+0.24} & 0.67 \ppl{+15\%} & \dm{-0.56} & 0.75 \nps{-25\%} & \dm{-0.94} & 0.48 \npl{-5\%} & \nn{-0.01} \\
\bottomrule
\end{tabular}%
}
\end{table*}

%% file: MainText/050_results.tex
\subsection{Main Results}
\label{sec:main_results}
We present findings from extensive experiments on \proposed.
Our goal is to evaluate how optimization strategies affect document visibility across the full generative search pipeline.
Existing research focuses on body text alone in SAGE, leaving the effectiveness of such a setting in realistic environments largely unexplored.
{Moreover, optimizing structural information (Section ~\ref{sec:structural_fields}), the process of organizing core document information into metadata fields so that search systems can correctly interpret the document, has not been studied in the context of generative search.}
We therefore separate optimization scope into three settings:
body text only, {structural information only,} and both,
enabling the first analysis of where gains and losses originate at each generative pipeline stage.

\medskip
\noindent\textbf{Limitation of Body Text only Optimization.}
As shown in Table~\ref{tab:main_results} (Left), optimizing body text alone consistently degrades visibility across all stages.
At retrieval, optimization strategies that replace common expressions with domain-specific terms (e.g., technical words) or uncommon vocabulary (e.g., unique words) show the largest retrieval drops. 
We attribute this to the lexical mismatch between optimized documents and user queries, which typically use common vocabulary.
For example, replacing terms like ``\textit{eating}'' with ``\textit{alimentary routines}'' or ``\textit{sleeping}'' with ``\textit{somnolence}'' directly reduces term overlap, causing BM25-based retrievers to assign lower relevance scores.
Notably, AutoGEO~\cite{wu2025generative} exhibits the largest degradation, with a retrieval rank drop of $-$22.35.
We find that AutoGEO tends to substantially expand the document content, introducing lengthy rewrites that dilute keyword density and shift the document further from the original query vocabulary.
At reranking, a similar but milder degradation is observed, suggesting that rewrites may introduce slight semantic shifts that rerankers are sensitive to.
However, even when the average rank drop is modest, small shifts at retrieval or reranking can be critical in practice.
In generative search, only top-ranked documents are passed to the generator, and a document falling below this threshold is excluded entirely, making it invisible at the generation stage regardless of its content quality.
At generation, gains remain relatively marginal across strategies, aligning with recent findings that presentational modifications offer limited visibility improvements~\cite{puerto2025c}.
Overall, body-only optimization is insufficient for enhancing visibility throughout the pipeline.
These findings suggest that effective optimization must extend beyond body text to structural information, which plays an important role in document ranking in real-world generative search systems.

\medskip
\noindent\textbf{Effectiveness of Structural Information Optimization.}
As shown in Table~\ref{tab:main_results} (Center), extending optimization scope to structural information significantly improves visibility across all strategies compared to body-only optimization.
The improvement is particularly notable at retrieval, with a +22\% boost in Hit Rate and a +2.72 average retrieval rank gain.
Structural information is inherently designed to be dense with query-relevant terms, increasing lexical overlap with user queries that BM25-based retrievers directly prioritize.
Accordingly, strategies that enrich content with keywords, entities, and numbers show strong effectiveness.
For example,  adding statistics rewrites a verbose meta description into a concise summary with specific facts like ``\textit{1983, comedy, Rotten Tomatoes, grossed $61M$},'' while adding quotations refines a generic title ``\textit{Panel Clarifies Advice}'' into ``\textit{IOM Panel Clarifies Vitamin D Guidance},'' introducing entity-specific terms that better align with user queries.
Interestingly, optimizing structural information alone also improves visibility at reranking and generation.
We attribute this partly to the increased number of documents that pass through retrieval into downstream stages, raising their chances of being reranked favorably and cited.
Additionally, optimizing structural information effectively places well-organized, informative summaries at the top of the document, consistent with recent findings that such positioning benefits document visibility in the generation stage~\cite{puerto2025c}.
However, applying optimization to both scopes (Table~\ref{tab:main_results} Right) yields lower visibility gains.
The negative retrieval effects of body-text modification, observed in the body-only setting, could partially diminish the gains from optimizing structural information.

\input{Figures/050_reranking_casestudy_fig}
\medskip
\noindent\textbf{Reranking as a Persistent Bottleneck.}
\label{sec:reranking_bottleneck}
All optimization strategies struggle at the reranking stage, showing consistent visibility degradation regardless of optimization scope (Table~\ref{tab:main_results}).
One contributing factor is that top-ranked documents are very close in relevance, making them more sensitive to rank displacement
from even minor content changes.
Although the average rank drop remains within 1 position across most settings,
even such small shifts can alter the relative ordering among closely ranked candidates.
Notably, 5.8\% of target documents in our experiments dropped from rank 10 to 11 during reranking, narrowly missing the input threshold for the generator in \proposed.
This highlights the importance of maintaining or improving rank position within generative search pipelines, as they inherently impose such cutoffs.
To better understand what factors positively or negatively affect reranking results, we conduct a case study analyzing documents that gained or lost rank after optimization.
Figure~\ref{fig:reranking_casestudy} presents two representative cases.
{It is worth noting that SAGEO optimizes documents without access to the incoming query, meaning content modifications cannot be tailored to specific query intents. Despite this constraint, clear patterns emerge in how the reranker responds to different types of optimizations.}
First, the reranker favors content additions that enhance alignment with the query's informational need, while penalizing additions that expand the document's scope beyond what the query seeks.
Second, placing the answer early in the document yields higher reranking scores, whereas restructuring that displaces the answer to later paragraphs results in significant rank drops, even when the answer itself remains intact. 
These observations indicate that reranking-aware optimization should preserve topical alignment and answer prominence over broad content expansion.

\input{Figures/050_domain_heatmap}
\subsection{In-depth Analysis}
To further understand the effectiveness of SAGEO, we analyze three dimensions: domain, citation source, and SAGEO backbone model.

\medskip
\noindent\textbf{Domain Analysis.}
Figure~\ref{fig:domain_heatmap_fig} shows the effectiveness of each optimization method across nine domains, reporting change in citation rate at the generation stage.
We observe that domains with general information-seeking queries (e.g., Web Search, General QA) benefit the most, as strategies like adding statistics provide concrete evidence that directly address these queries.
Shopping is the only domain where every optimization method decreases citation likelihood.
Product documents are often already well-organized for their customers, and the queries in this domain are predominantly casual and recommendation-seeking (e.g., ``\textit{gift ideas for my friend}''). 
Optimizing such documents may shift them away from this expected tone, reducing their competitiveness in the generator's response.
These results suggest that optimization strategies applied without considering domain-specific user intent can be ineffective or even harmful, emphasizing the need for domain-aware optimization.

\medskip
\noindent\textbf{Citation Source Analysis.}
To understand what drives citation at the generation stage, we examine how generative models interact with document content when forming responses.
To achieve this, we conduct an experiment prompting the generator to provide the exact quote used for each citation in its response.
We then apply fuzzy matching to locate each quoted segment in the source document and identify the region from which it originates.
Figure~\ref{fig:citation_position} illustrates the results as a density plot.
Each colored region corresponds to a structural component (e.g., title, meta description, headings, JSON-LD) or body text, and values above the horizontal baseline ($y=1.0$) indicate that the region tends to be cited more frequently.
We observe that the vast majority of citations originate from body text, while structural information is cited less frequently despite its strong contribution to retrieval.
We attribute this to the information density of body text, which provides richer evidence for addressing user queries compared to the concise, keyword-oriented nature of structural information.
Combined with the earlier finding that structural information improves visibility at the generation stage (Section~\ref{sec:main_results}), these results suggest that structural information plays an important role in surfacing documents, but the generator primarily references body text when forming responses.
Thus, structural information and body text serve complementary roles in SAGE, and effective optimization requires addressing both.

\input{Figures/050_position_density_overall_}
\input{Figures/050_opensource_sageo_fig}
\medskip \noindent\textbf{{SAGEO Backbone Model Analysis.}} 
To examine whether {the choice of backbone model for SAGEO} exhibits distinct optimization behaviors, we conduct experiments with two additional open-source models, LLaMA-3.3-70B and Qwen3-80B (Figure~\ref{fig:opensource_sageo}).
We compare them against GPT-5-mini, {applying the All-in-One strategy across all three models.}
Interestingly, LLaMA-3.3-70B achieves the highest win rate (42.2\%) at retrieval, surpassing GPT-5-mini (38.0\%).
We find that LLaMA often generates short, keyword-dense text (296 words on average vs.\ 782 for GPT).
Because BM25 rewards keyword matches and applies length normalization that favors shorter documents, this style is naturally advantaged at the retrieval stage.
However, this advantage reverses at reranking, where GPT-5-mini dominates with a 63.1\% win rate while LLaMA-3.3-70B drops to 13.0\%. 
Unlike BM25, the LLM-based reranker evaluates both relevance and coherence of the documents to the user query, and tends to penalize irrelevant content~\cite{shi2024replug,ke2024bridging}.
A case study reveals that LLaMA-optimized documents frequently mention query terms without meaningfully addressing the underlying question.
For instance, given the query ``\textit{What are some of the most useful vim shortcuts?}'', LLaMA opens with an unrelated introduction while GPT directly lists shortcuts with clear formatting.
At generation, the gap widens further. 
GPT-5-mini reaches a 73.2\% win rate while LLaMA-3.3-70B falls to 7.0\%.
These results demonstrate that short, keyword-dense documents can succeed at retrieval, but downstream stages increasingly reward content that substantively addresses the query.

\input{Tables/070_our_method_table}
\subsection{Insights \& Exploration}
\label{subsec:insights}
In this section, we study effective strategy combinations and introduce practical optimization guidelines informed by our findings.

\medskip
\noindent\textbf{Strategy Combination.}
When applying SAGEO, multiple strategies can be combined.
However, the All-in-One optimization results in Table~\ref{tab:main_results} suggest that combining all strategies at once is not always effective.
We therefore selectively evaluate combinations of top-performing strategies (Table~\ref{tab:our_method_table}).
While these combinations improve generation performance, retrieval rank degradation persists across all pairs.
This indicates that naive pairing cannot simultaneously benefit all pipeline stages, motivating a stage-targeted approach.

\medskip \noindent\textbf{Practical Guidelines.} 
To this end, we introduce stage-aware optimization, which tailors optimization to the specific priorities of each pipeline stage.
Our method builds on the following principles:
\begin{itemize}
    \item \textbf{Enrich structural fields with key entities.} Core entities, numbers, and terms from body text are organized in structural fields to strengthen query-document matching at retrieval.
    \item \textbf{Make claims prominent and self-contained.} The main claim is placed at the start of the body, and each statement is ensured to carry specific evidence, increasing citability at generation.
    \item \textbf{Maintain topical coherence across paragraphs.} Ambiguous pronouns are replaced with explicit subject references, and core terms are naturally emphasized to keep paragraphs connected.
    \item \textbf{Adapt to domain context.} The document's domain is assessed before any changes, and both domain characteristics and user intent are considered to determine how to optimize. 
\end{itemize}
We demonstrate the effectiveness of our method in Table~\ref{tab:our_method_table}, where it achieves competitive performance across all stages, with notable gains at both reranking and generation among all strategies evaluated.
These results suggest that \proposed serves as an effective evaluation environment for developing practical SAGEO methods, enabling stage-level analysis and providing actionable insights.
The detailed prompt used in our method is provided in Appendix~\ref{app:prompt}.

\input{Tables/053_retriever_comparison_table}
\input{Tables/054_reranker_comparison_table}
\subsection{Supplementary Analysis}
The primary objective of SAGEO Arena is to derive actionable findings that generalize to real-world SAGE.
We therefore adopt BM25 as the first-stage retriever, the standard choice for web-scale retrieval given its scalability and efficiency, paired with representative models for reranking and generation.
Nevertheless, we verify the robustness of our findings with alternative model choices, varying the retriever, reranker, and generator in the body-only setting.

\medskip
\noindent\textbf{Robustness across Retrievers.}
We additionally evaluate two alternative settings: BGE-base-en-v1.5~\cite{bge_embedding} as a dense retriever, and its combination with BM25 as a hybrid retriever.
As shown in Table 4, the trend is consistent across all three settings. 
Body-only optimization does not reliably improve retrieval visibility, and often lowers the target document's rank. 
This confirms that the observed degradation is not specific to BM25, but reflects a broader limitation of optimizing body text alone regardless of the underlying retriever.

\medskip
\noindent\textbf{Robustness across Rerankers and Generators.}
We further test two substitutions: replacing the reranker with BGE-Reranker-v2-m3~\cite{li2023making,chen2024bge} and replacing the generator with Claude-Sonnet-4.6. 
As shown in Table 5, the trend persists in both cases.
Specifically, the reranker substitution moderately amplifies the degradation, indicating that lighter rerankers are more sensitive to body-only optimization.
The generator substitution also worsens both Citation Rate and Rank Change for most strategies.
Overall, these results confirm that body-only optimization can degrade visibility across the full pipeline, motivating stage-aware optimization in SAGE.

%% file: Figures/050_reranking_casestudy_fig.tex
\begin{figure}[t]
\centering
\includegraphics[width=0.93\linewidth]{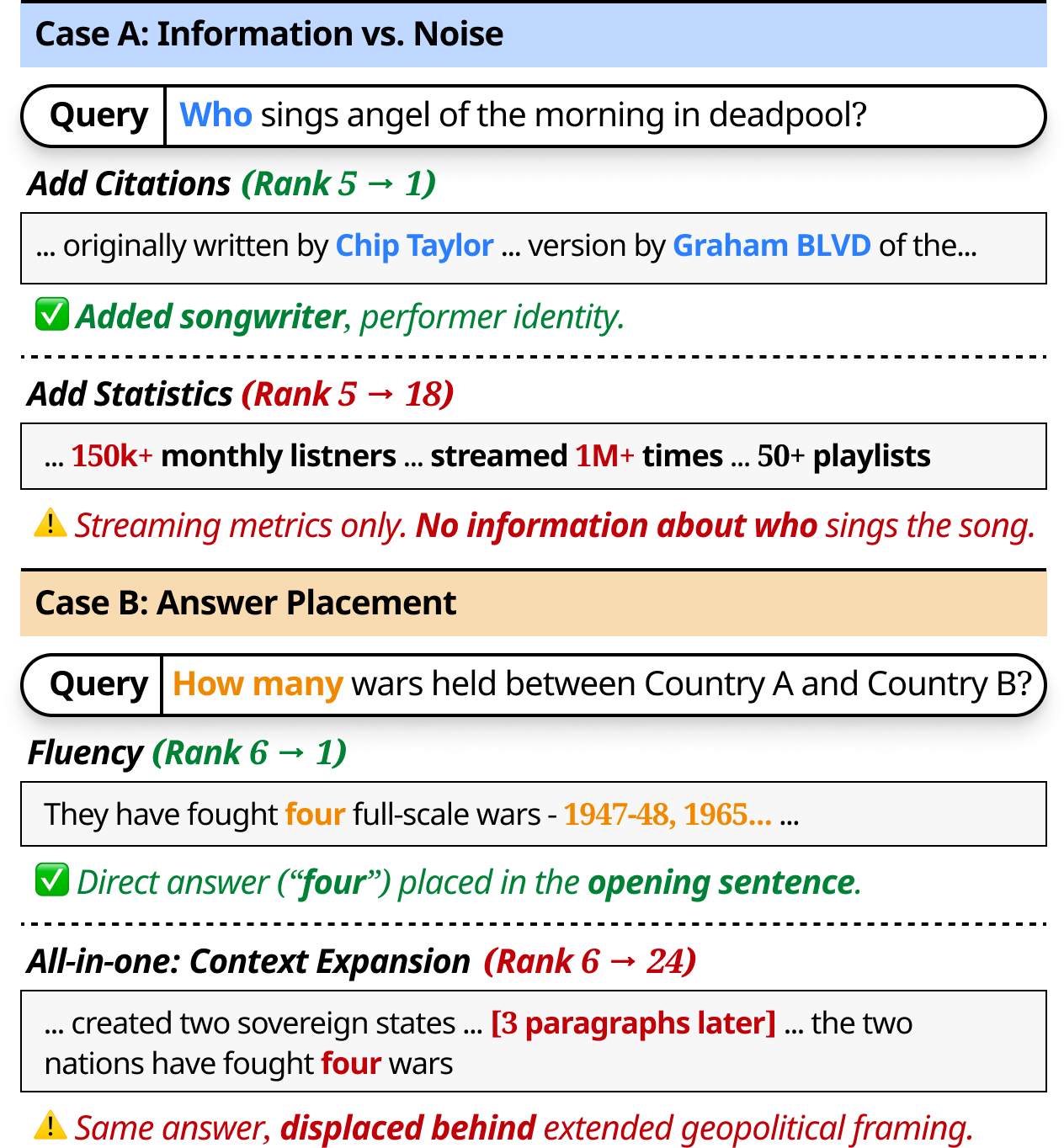}
\caption{
Reranking case studies illustrating two factors that improve document ranking after optimization. Case A shows that adding content that directly addresses the query's informational need boosts rank. Case B shows that placing the  direct answer in early paragraphs improves ranking position.
}
\label{fig:reranking_casestudy}
\vspace{-0.32cm} 
\end{figure}

%% file: Figures/050_domain_heatmap.tex
\begin{figure}[t]
\centering
\includegraphics[width=0.99\linewidth]{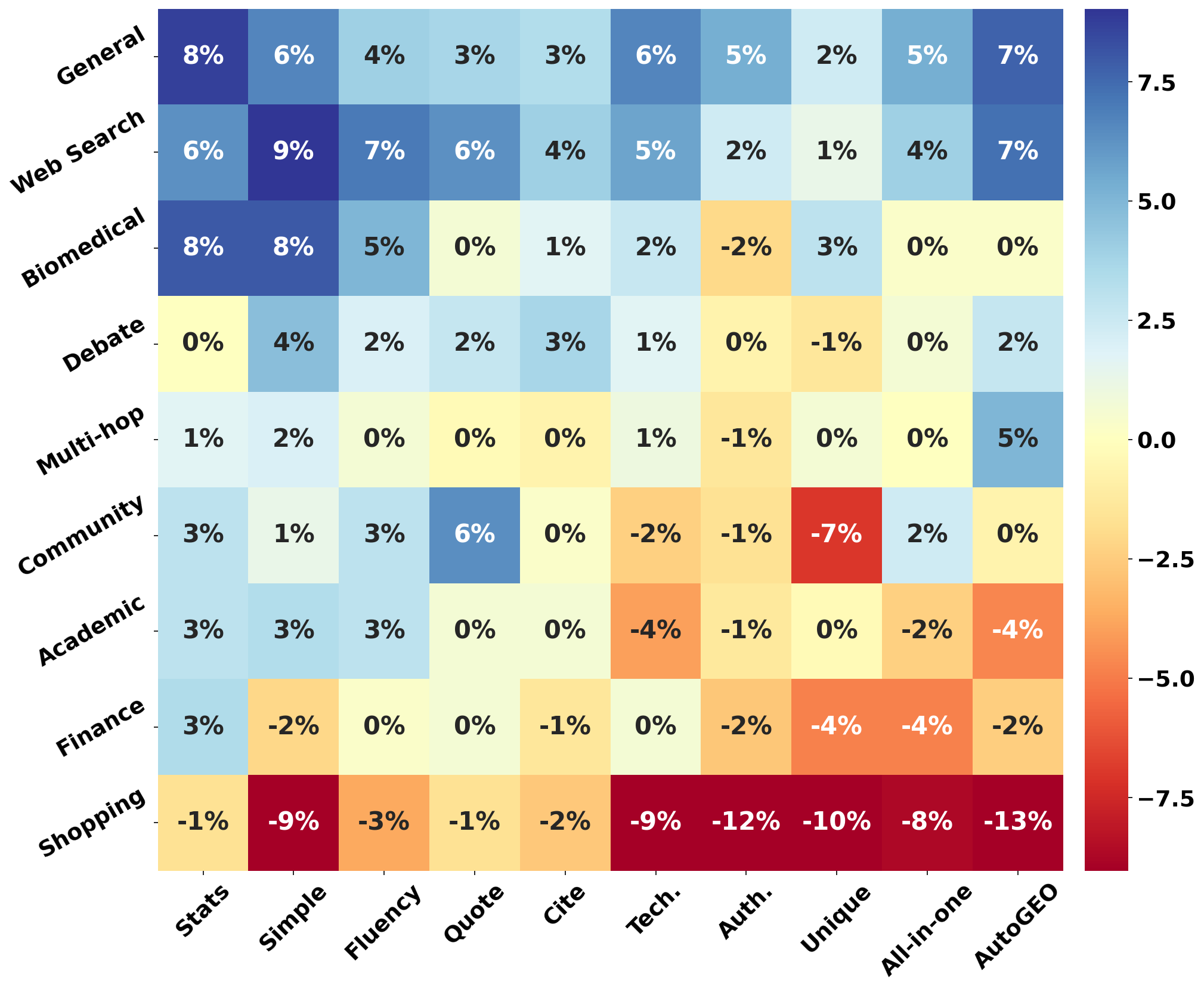}
\caption{
Effectiveness of each optimization strategy across nine domains, measured by change in citation rate at the generation stage. Positive \% indicate increased citation rate.
}
\label{fig:domain_heatmap_fig} 
\end{figure}

%% file: Figures/050_position_density_overall_.tex
\begin{figure}[t]
\centering
\includegraphics[width=0.91\linewidth]{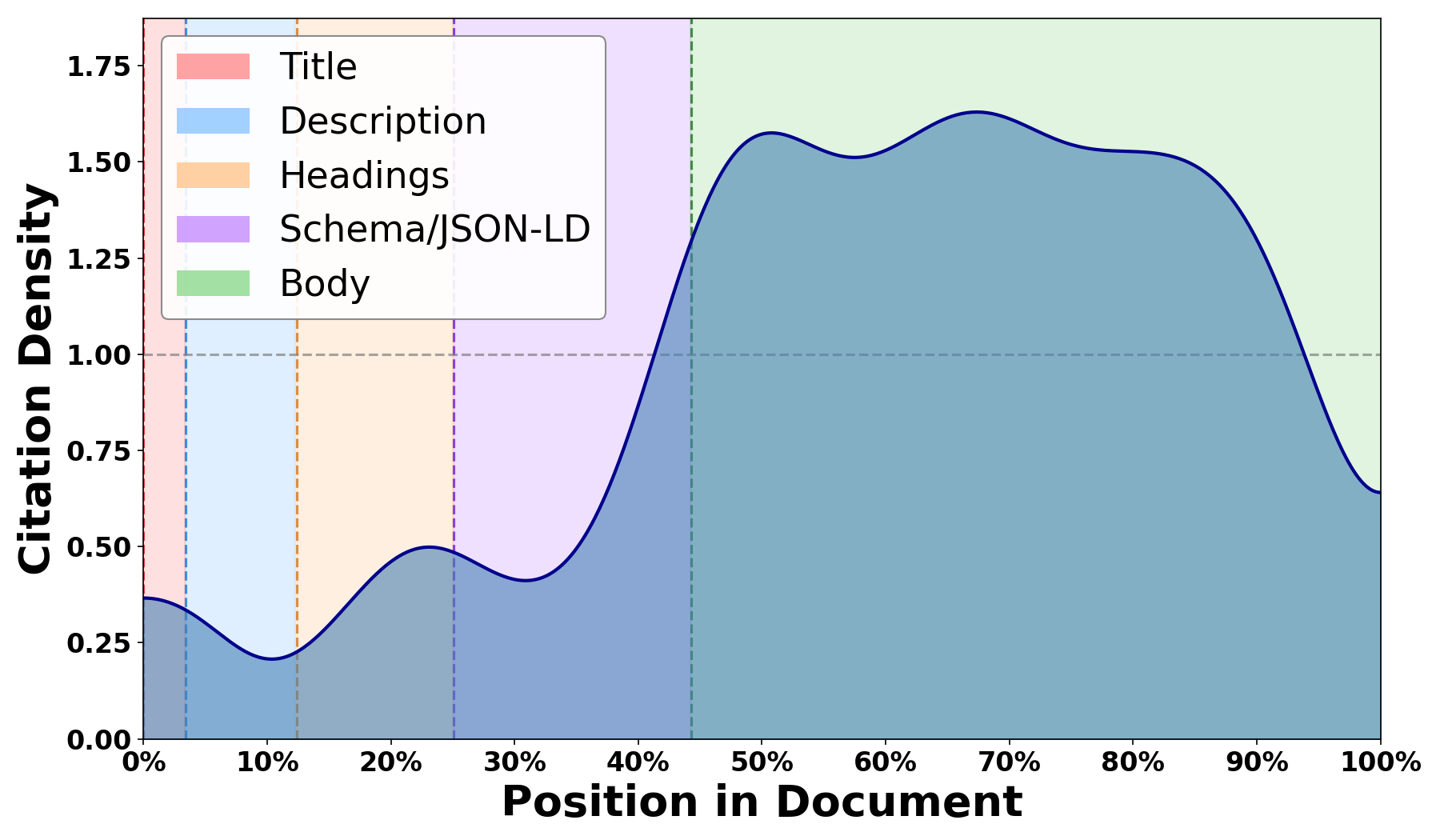}
\caption{
Density of citation sources across document regions. 
}
\label{fig:citation_position}
\vspace{-0.3cm} 
\end{figure}

%% file: Figures/050_opensource_sageo_fig.tex
\begin{figure}[ht]
\centering
\includegraphics[width=0.99\linewidth]{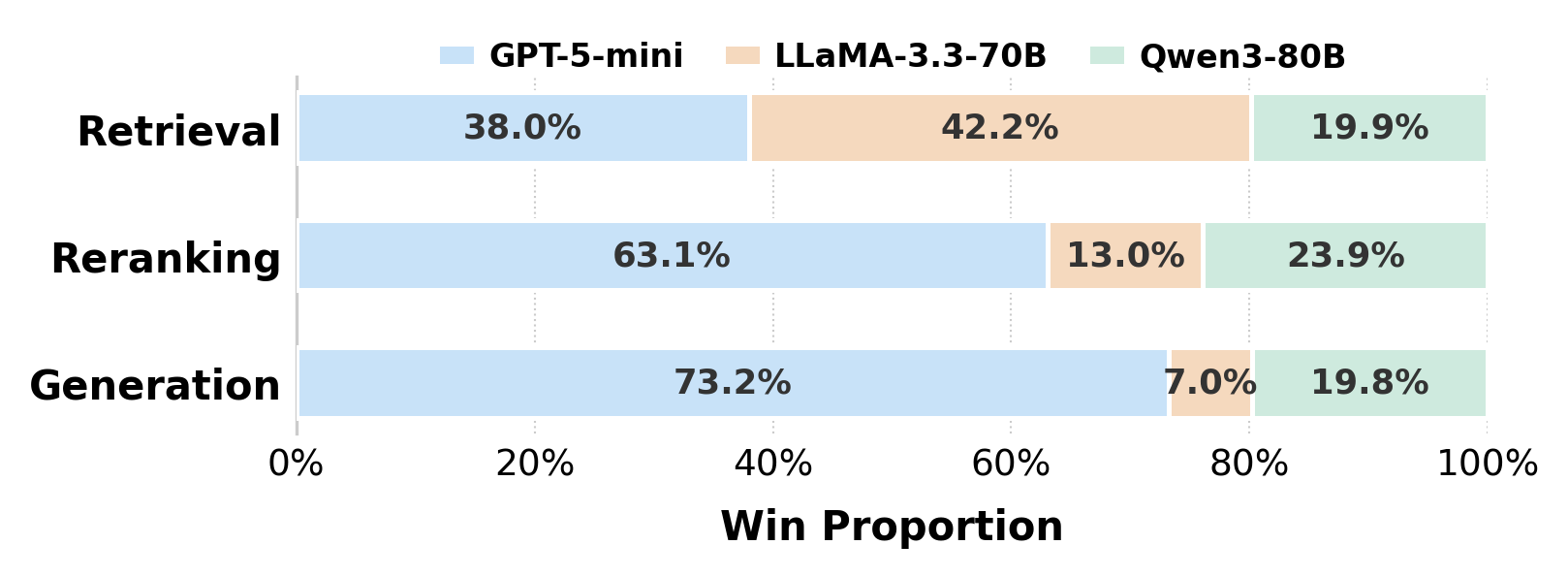}
\caption{
Win rate comparison across three backbone models (GPT-5-mini, LLaMA-3.3-70B, Qwen3-80B) at each stage. A model is considered a win when its optimized document achieves the highest rank among the three at a given stage.
}
\label{fig:opensource_sageo}
\end{figure}

%% file: Tables/070_our_method_table.tex
\begin{table}[t]
\caption{Comparison of SAGEO strategy combination and \textsc{StageAware} optimization (Ours) under the Both setting. \textsc{StageAware} achieves the strongest SAGEO performance.}
\centering
\label{tab:our_method_table}
\setlength{\tabcolsep}{2pt}
\renewcommand{\arraystretch}{1.15}
\footnotesize
\resizebox{0.9\columnwidth}{!}{%
\begin{tabular}{l cc cc cc}
\toprule
\multirow{2}{*}{\textbf{Method}} 
& \multicolumn{2}{c}{Retrieval} & \multicolumn{2}{c}{Reranking} & \multicolumn{2}{c}{Generation} \\
\cmidrule(lr){2-3} \cmidrule(lr){4-5} \cmidrule(lr){6-7}
& H@20 & $\Delta$Rank & H@10 & $\Delta$Rank & Cite & $\Delta$Rank \\
\midrule
\rowcolor{gray!5}
\textit{Baseline} & 0.58 & \nn{-} & 1.00 & \nn{-} & 0.50 & \nn{-} \\
\midrule
\rowcolor{gray!10}
\rowcolor{gray!10}
\multicolumn{7}{l}{\textit{\textbf{Combined Strategy}}} \\
EasyLang + Quote      & 0.67 \ppl{+14\%} & \dm{-0.25}  & 0.78 & \dm{-0.33} & 0.54 & \gl{+0.64} \\
EasyLang + Stats      & 0.65 \ppl{+11\%} & \dm{-1.77}  & 0.76 & \dm{-0.56} & 0.51 & \gl{+0.38} \\
Fluency + Quote   & 0.66 \ppl{+13\%} & \dm{-0.63}  & 0.77 & \dm{-0.43} & 0.54 & \gl{+0.65} \\
Fluency + Stats   & 0.65 \ppl{+11\%} & \dm{-1.63}  & 0.76 & \dm{-0.55} & 0.53 & \gl{+0.60} \\
\midrule
\textbf{\textsc{StageAware}\xspace(Ours)} &\textbf{0.75} \ppm{\textbf{+28}\%} & \gm{\textbf{+4.86}} & \textbf{0.80} & \nn{\textbf{-0.08}} & \textbf{0.58} & \gl{\textbf{+1.01}} \\
\bottomrule
\end{tabular}%
}
\end{table}

%% file: Tables/053_retriever_comparison_table.tex
\begin{table}[t]
  \caption{Effectiveness of SAGEO strategies across different retrievers under the body-only setting. We report Hit Rate (H@20) and $\Delta$Rank for BM25, Dense, and Hybrid retrievers.}
  \centering
  \label{tab:retriever_variation}
  \setlength{\tabcolsep}{2pt}
  \renewcommand{\arraystretch}{1.15}
  \footnotesize
  \resizebox{0.94\columnwidth}{!}{%
  \begin{tabular}{l cc cc cc}
  \toprule
  \multirow{2}{*}{\textbf{Strategy}}
  & \multicolumn{2}{c}{\textbf{BM25}}
  & \multicolumn{2}{c}{\textbf{Dense}}
  & \multicolumn{2}{c}{\textbf{Hybrid}} \\
  \cmidrule(lr){2-3} \cmidrule(lr){4-5} \cmidrule(lr){6-7}
  & H@20 & $\Delta$Rank & H@20 & $\Delta$Rank & H@20 & $\Delta$Rank \\
  \midrule
  \rowcolor{gray!5}
  \textit{Baseline} & 0.58 \bp & \nn{-} & 0.68 \bp & \nn{-} & 0.67 \bp & \nn{-} \\
  \midrule
  Auth.       & 0.57 \npl{-1\%}  & \dm{-0.20}  & 0.67 \npl{-1\%} & \nn{+0.10} & 0.66 \npl{-1\%}  & \dm{-0.61} \\
  Cite        & 0.56 \npl{-3\%}  & \dm{-1.50}  & 0.66 \npl{-3\%} & \dm{-1.01} & 0.65 \npl{-4\%} & \dd{-2.80} \\
  EasyLang    & 0.52 \npl{-10\%} & \dd{-4.18}  & 0.65 \npl{-4\%} & \dd{-2.20} & 0.62 \npl{-8\%} & \dd{-4.32} \\
  Fluency     & 0.57 \npl{-1\%}  & \dm{-0.71}  & 0.66 \npl{-2\%} & \dm{-0.46} & 0.65 \npl{-3\%} & \dm{-1.42} \\
  Quote       & 0.57 \npl{-1\%}  & \dm{-0.33}  & 0.66 \npl{-2\%} & \dm{-1.05} & 0.65 \npl{-2\%} & \dm{-1.68} \\
  Stats       & 0.57 \npl{-2\%}  & \dm{-0.51}  & 0.68 \ppl{+1\%}  & \nn{+0.76} & 0.66 \npl{-2\%} & \dm{-0.98} \\
  Tech.       & 0.50 \npm{-14\%} & \ds{-6.23}  & 0.65 \npl{-4\%} & \dm{-1.85} & 0.60 \npl{-11\%} & \ds{-5.31} \\
  Unique      & 0.53 \npl{-8\%}  & \dd{-3.47}  & 0.66 \npl{-3\%} & \dm{-1.07} & 0.63 \npl{-7\%} & \dd{-3.20} \\
  All-in-One  & 0.50 \npm{-14\%} & \ds{-5.93}  & 0.66 \npl{-2\%} & \dm{-1.26} & 0.62 \npl{-8\%} & \dd{-4.25} \\
  AutoGEO     & 0.37 \nps{-36\%} & \ds{-22.35} & 0.66 \npl{-2\%} & \dm{-1.49} & 0.63 \npl{-7\%} & \dd{-3.57} \\
  \midrule
  \textbf{Avg.} & 0.53 \npl{-9\%} & \dd{-4.54} & 0.66 \npl{-2\%} & \dm{-0.95} & 0.64 \npl{-5\%} & \dd{-2.81} \\
  \bottomrule
  \end{tabular}%
  }
\vspace{-0.2cm} 
\end{table}

%% file: Tables/054_reranker_comparison_table.tex
\begin{table}[t]
  \caption{Results of SAGEO strategies under BM25 retrieval with alternative rerankers or generators in the body-only setting. Each evaluation varies one component independently, with the default model shown on the left within each block.}
  \centering
  \label{tab:reranker_generator_results}
  \setlength{\tabcolsep}{2pt}
  \renewcommand{\arraystretch}{1.15}
  \footnotesize
  \resizebox{0.99\columnwidth}{!}{%
  \begin{tabular}{l cc cc | cc cc}
  \toprule
  \multirow{3}{*}[-0.5em]{\textbf{Strategy}}
  & \multicolumn{4}{c|}{\textbf{Reranker}}
  & \multicolumn{4}{c}{\textbf{Generator}} \\
  \cmidrule(lr){2-5} \cmidrule(lr){6-9}
  & \multicolumn{2}{c}{Qwen3-Reranker}
  & \multicolumn{2}{c|}{BGE-Reranker-v2-m3}
  & \multicolumn{2}{c}{GPT-5-mini}
  & \multicolumn{2}{c}{Claude Sonnet} \\
  \cmidrule(lr){2-3} \cmidrule(lr){4-5}
  \cmidrule(lr){6-7} \cmidrule(lr){8-9}
  & H@10 & $\Delta$Rank & H@10 & $\Delta$Rank
  & Cite & $\Delta$Rank & Cite & $\Delta$Rank \\
  \midrule
  \rowcolor{gray!5}
  \textit{Baseline} & 1.00 \bp & \nn{-} & 1.00 \bp & \nn{-} & 0.50 \bp & \nn{-} & 0.77 \bp & \nn{-} \\
  \midrule
  Auth.       & 0.91 \npl{-9\%}  & \dm{-0.24} & 0.91 \npl{-9\%}  & \dm{-0.44} & 0.49 \npl{-3\%}  & \dm{-0.10} & 0.75 \npl{-3\%}  & \dm{-0.11} \\
  Cite        & 0.87 \npm{-13\%} & \dm{-0.45} & 0.85 \npm{-15\%} & \dm{-0.94} & 0.48 \npl{-4\%}  & \dm{-0.13} & 0.69 \npl{-11\%} & \dm{-0.51} \\
  EasyLang    & 0.84 \npm{-16\%} & \dm{-0.54} & 0.78 \npm{-22\%} & \dm{-1.45} & 0.49 \npl{-4\%}  & \dm{-0.09} & 0.67 \npm{-13\%} & \dm{-0.78} \\
  Fluency     & 0.91 \npl{-9\%}  & \dm{-0.18} & 0.92 \npl{-8\%}  & \dm{-0.31} & 0.50 \npl{-1\%}  & \nn{-0.01} & 0.74 \npl{-4\%}  & \dm{-0.14} \\
  Quote       & 0.90 \npl{-10\%} & \dm{-0.36} & 0.89 \npl{-11\%} & \dm{-0.69} & 0.47 \npl{-7\%}  & \dm{-0.26} & 0.71 \npl{-8\%}  & \dm{-0.45} \\
  Stats       & 0.90 \npl{-10\%} & \dm{-0.33} & 0.90 \npl{-10\%} & \dm{-0.54} & 0.48 \npl{-4\%}  & \dm{-0.18} & 0.76 \npl{-1\%}  & \nn{+0.03} \\
  Tech.       & 0.80 \npm{-20\%} & \dm{-1.03} & 0.69 \nps{-31\%} & \dd{-2.08} & 0.47 \npl{-6\%}  & \dm{-0.15} & 0.67 \npm{-13\%} & \dm{-0.80} \\
  Unique      & 0.86 \npm{-14\%} & \dm{-0.70} & 0.81 \npm{-19\%} & \dm{-1.33} & 0.46 \npl{-8\%}  & \dm{-0.33} & 0.68 \npl{-12\%} & \dm{-0.80} \\
  All-in-One  & 0.83 \npm{-17\%} & \dm{-0.68} & 0.73 \nps{-27\%} & \dm{-1.83} & 0.49 \npl{-2\%}  & \nn{-0.03} & 0.69 \npl{-11\%} & \dm{-0.60} \\
  AutoGEO     & 0.58 \nps{-42\%} & \dd{-2.28} & 0.65 \nps{-35\%} & \dd{-2.17} & 0.39 \npm{-22\%} & \dm{-0.78} & 0.48 \nps{-38\%} & \dm{-1.94} \\
  \midrule
  \textbf{Avg.} & 0.84 \npm{-16\%} & \dm{-0.68} & 0.81 \npm{-19\%} & \dm{-1.18} & 0.47 \npl{-6\%} & \dm{-0.21} & 0.68 \npl{-12\%} & \dm{-0.61} \\
  \bottomrule
  \end{tabular}%
  }
\vspace{-0.24cm} 
\end{table}

%% file: MainText/060_conclusion.tex
In this paper, we introduce \proposed, a realistic evaluation environment for stage-level visibility analysis of Search-Augmented Generative Engine Optimization. 
By integrating a full generative search pipeline with a large-scale corpus preserving structural information, \proposed enables investigation of how optimization signals propagate in generative search.
We observe that findings from existing benchmarks do not readily generalize to realistic settings, and that stage-aware optimization targeting each pipeline stage is crucial for visibility in generative search.
Overall, \proposed offers a reproducible environment for developing optimization strategies that generalize to real-world generative search.

%% file: MainText/061_acknowledgements.tex
This work was supported by the IITP grants funded by the Korea government (MSIT) (RS-2024-00457882, AI Research Hub Project; RS-2026-25520654).

%% file: MainText/070_appendix.tex
\section{Benchmark Construction}
\label{appendix:dataset}
\medskip
\noindent\textbf{Statistics.}
SAGEO Arena comprises nine domains sourced from established query datasets: MS MARCO~\cite{bajaj2018msmarcohumangenerated} (Web Search), Natural Questions (General QA)~\cite{kwiatkowski-etal-2019-natural}, HotpotQA~\cite{yang2018hotpotqa} (Multi-hop QA), NFCorpus~\cite{Boteva2016Nfcorpus} (Biomedical), Quora~\cite{sharma2019naturallanguageunderstandingquora} (Community QA), FiQA~\cite{maia2018www} (Finance), DebateQA~\cite{xu2024debateqaevaluatingquestionanswering} (Debate), E-commerce (Shopping), and Researchy (Academic)~\cite{wu2025generative}. 
Domain selection ensures diversity in content characteristics, query complexity, and information-seeking behaviors. 
Detailed statistics are presented in Table \ref{tab:benchmark_stats}.
\input{Tables/030_benchmark_stats_table}

\medskip
\noindent\textbf{Difference from RAG Benchmarks.}
Standard RAG benchmarks evaluate whether a system generates faithful and relevant answers given a query and a document corpus.
While this setting is useful for measuring answer quality, it does not directly capture the content creator’s perspective, where the central question is whether a specific document can remain visible across retrieval, reranking, and generation.
SAGEO Arena addresses this gap by approximating the real-world generative search pipeline and measuring target document visibility before and after optimization.
Moreover, SAGEO Arena preserves structural information from web documents.
These fields provide useful signals for how web documents are interpreted and surfaced in search systems, but are typically discarded in RAG-style benchmarks that operate mainly on plain body text passages.

\section{Implementation Details}
\label{appendix:implementation_details}
We instantiate the pipeline described in Section~\ref{sec:pipeline} with the following configuration.
We segment document body text into passages of 256 tokens with a 64-token overlap.
During indexing, we assign equal weights across structural fields and chunked passages to establish a controlled and fair baseline.
In practice, search engines apply varying weights to different fields~\cite{elastic-app-search-relevance-tuning};
however, adopting uniform weights ensures that observed visibility changes stem from optimization strategies rather than field-specific biases.
We retrieve the top-$100$ passages per query and rescore them using Qwen3-Reranker-4B~\cite{qwen3embedding}.
For generation, the top-$10$ candidates are retained for response synthesis, using GPT-5-mini as the default generation model.
For each test query, we first establish a baseline by
running the query through the full search pipeline,
then randomly select a target document from the top-$10$
candidates that enter the generation stage (i.e., ranked
within the top-$10$ after reranking), ensuring sufficient
relevance to the test query.
All SAGEO optimization strategies are implemented using GPT-5-mini with strategy-specific prompts following~\cite{aggarwal2024geo} for the eight content modification strategies, \cite{puerto2025c} for All-in-One, and~\cite{wu2025generative} for AutoGEO.

\input{Figures/050_query_robustness_fig}
\input{Tables/071_our_method_prompt}
\section{Further Analysis}
\medskip
\noindent\textbf{Optimization strategies show limited robustness across query formulations.}
Real users express information needs through diverse query formulations, making robustness to query variation essential for effective optimization.
To examine this, we evaluate optimized documents against four query rewriting strategies: expansion (adding specificity), simplification (keeping core keywords), rephrasing (lexical substitution), and abstraction (semantic generalization).
We report results averaged across eight optimization strategies from
~\cite{aggarwal2024geo} in Figure~\ref{fig:query_robustness_fig}.
We observe consistent visibility degradation under all query variations, with a progressive drop as queries deviate further from the original phrasing.
Since document optimization is performed without knowledge of the incoming query, we find that it tends to elaborate on the document's existing content, making the document more specific to its original topic and vocabulary. 
As a result, optimization effectiveness becomes sensitive to query formulation, especially when rewritten queries reduce lexical overlap with the optimized document. 
For example, the degradation is most pronounced under abstraction, where queries are generalized beyond the document’s original wording, causing the most severe drop.
These results indicate that current strategies show limited transferability, motivating robust SAGEO methods that preserve visibility across diverse query formulations.

\medskip
\noindent\textbf{Answer Quality Analysis.}
Although our primary focus is to evaluate document visibility, we provide additional analysis on SAGEO's impact on answer quality.
SAGEO Arena studies white-hat document optimization, where methods cooperatively enhance content rather than inject adversarial signals. 
Moreover, each generated answer is grounded in multiple retrieved documents, so modifying one target document is unlikely to meaningfully change overall answer quality.
To empirically validate this, we conduct a sanity check using 3,000 before-and-after optimization samples, evaluating whether generated answers address the query's information need using GPT-5.4 as the evaluator.
We observe only a small change after optimization across all methods, from 88.9\% to 87.3\%.

\section{Limitations}
While SAGEO Arena approximates a realistic generative search pipeline to provide actionable insights, it does not fully reproduce commercial search engines with proprietary ranking signals or user behavior signals.
Since these signals are difficult to access and reproduce, SAGEO Arena focuses on controllable factors that content creators can observe and directly modify to improve document visibility.
Nevertheless, incorporating richer ranking and user behavior signals remains an important direction for future SAGEO research.

\section{Prompt for StageAware SAGEO}
\label{app:prompt}
Based on the empirical findings from SAGEO Arena, we introduce a stage-aware SAGEO strategy that tailors optimization to the specific priorities of each pipeline stage. 
The prompt  is shown in Table \ref{tab:our_method_prompt}.

%% file: Tables/030_benchmark_stats_table.tex
\begin{table}[ht]
\centering
\caption{Benchmark statistics across nine domains.}
\label{tab:benchmark_stats}
\small
\resizebox{0.97\columnwidth}{!}{%
\begin{tabular}{llcc}
\toprule
\textbf{Source Dataset} & \textbf{Domain} & \textbf{\#Sampled Queries} & \textbf{\#Retrieved Docs} \\
\midrule
MS MARCO            & Web Search    & 300 & 21,880 \\
Natural Questions   & General QA    & 300 & 21,921 \\
HotpotQA            & Multi-hop QA  & 300 & 13,409 \\
NFCorpus            & Biomedical    & 300 & 21,079 \\
Quora               & Community QA  & 300 & 21,734 \\
FiQA                & Finance       & 300 & 16,771 \\
DebateQA            & Debate        & 300 & 17,443 \\
E-commerce          & Shopping      & 300 & 18,885 \\
Researchy           & Academic      & 300 & 17,881 \\
\midrule
\textbf{Total}      &               & \textbf{2,700} & \textbf{171,003} \\
\bottomrule
\end{tabular}%
}
\end{table}

%% file: Figures/050_query_robustness_fig.tex
\begin{figure}[ht]
\centering
\includegraphics[width=0.99\linewidth]{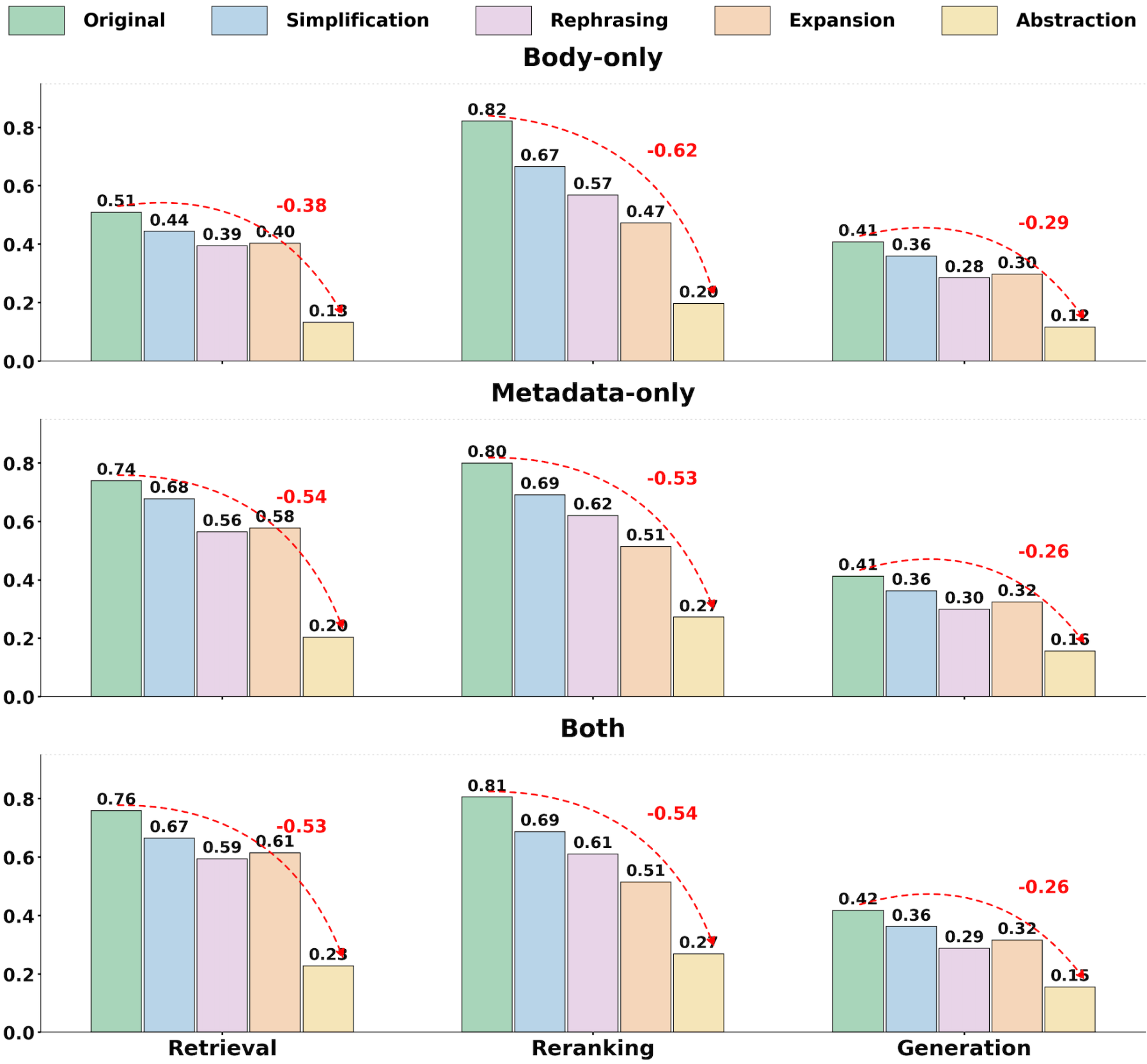}
\caption{
Document visibility across pipeline stages under four query variations for three optimization scopes. Values show mean Hit Rate at retrieval and reranking, and mean Citation Rate at generation, averaged across eight optimization strategies. Red numbers indicate percentage point change from the original query to the lowest-performing variation.
}
\label{fig:query_robustness_fig}
\end{figure}

%% file: Tables/071_our_method_prompt.tex
\begin{table*}[!h]
    \small
    \centering
    \caption{The prompt template for stage-aware SAGEO.}
    \begin{tabular}{p{16cm}}
    \toprule
    \textbf{Stage-Aware SAGEO Prompt} \\
    \midrule
    \textcolor{teal}{\textbf{[Task Description]}}\\
    Optimize the following document with these strategies. Keep the content faithful to the original.\\\\
    
    \textcolor{teal}{\textbf{[Pre-Optimization Considerations]}}\\
    \textbf{Before optimizing, think about two things:}\\
    1. \textbf{Domain}: Consider what domain this document belongs to (e.g., medical, finance, e-commerce, technical, casual). Match the tone and vocabulary to what readers in that domain expect. \\
    2. \textbf{Quality}: If a field is already clear, specific, and well-written, keep it as-is. Only optimize fields that genuinely benefit from it. Not every document needs heavy changes.\\\\
    
    \textcolor{teal}{\textbf{[Optimization Strategies]}}\\
    \textbf{1. Entity mirroring (structural fields):}\\
    Incorporate key entities, numbers, and domain terms from the body into the title, meta\_description, headings, and jsonld\_text. Add a keyword-rich summary sentence while keeping it compact. Skip if the structural fields already contain the right keywords.\\\\
    
    \textbf{2. Fluent, easy language (all text):}\\
    Rewrite sentences to be smooth, clear, and easy to read. Use simple words and short sentences. Avoid jargon when a plain alternative exists. If the writing is already clear and fluent, leave it unchanged.\\\\
    
    \textbf{3. Concrete evidence (body text):}\\
    Make claims specific. Bring front the main claim to the very start of the body. Each claim should be self-contained — a reader should understand it without reading surrounding text. If claims are already specific, do not rephrase them.\\\\
    
    \textbf{4. Keyword reinforcement (body text):}\\
    Naturally repeat the document's core topic terms and key phrases throughout the body. Use the main subject name instead of pronouns where it reads naturally. This keeps every paragraph clearly connected to the topic.\\
    \bottomrule
    \end{tabular}
    \label{tab:our_method_prompt}
\end{table*}